# Highly Catalytic Nanodots with Renal Clearance for Radiation Protection

*Xiao-Dong Zhang [1,3,] *, Jinxuan Zhang [2], Junying Wang [1], Jiang Yang [4], Jie Chen [3], Xiu Shen [3], Jiao Deng [5], Dehui Deng [5], Wei Long [3], Yuan-Ming Sun [3], Changlong Liu,[1] Meixian Li [2,] ***

[1] Department of Physics, School of Science, Tianjin University, Tianjin 300072, PR China

[2] Institute of Analytical Chemistry, College of Chemistry and Molecular Engineering, Peking University, Beijing 100871, China

[3] Tianjin Key Laboratory of Molecular Nuclear Medicine, Institute of Radiation Medicine, Chinese Academy of Medical Sciences and Peking Union Medical College, No. 238, Baidi Road, Tianjin 300192, China

[4] Environment, Energy and Natural Resources Center, Department of Environmental Science and Engineering, Fudan University, No.220, Handan Road, 200433, China

[5] State Key Laboratory of Catalysis, iChEM, Dalian Institute of Chemical Physics, Chinese Academy of Sciences, Dalian 116023, China

Correspondence should be addressed to X.Z. (xiaodongzhang@tju.edu.cn),

M.L.(lmwx@pku.edu.cn)

ABSTRACT: Ionizing radiation (gamma and x-ray) is widely used in industry and medicine, but it can also pose a significant hazardous effect on health and induce cancer, physical deformity and even death, due to DNA damages and invasion of free radicals. There is therefore an urgent unmet demand in designing highly efficient radioprotectants with synergetic integration of effective renal clearance and low toxicity. In this study, we designed ultrasmall (sub-5 nm) highly catalytically active and cysteine-protected $MoS_2$ dots as radioprotectants and investigated their application in protection against ionizing radiation. *In vivo* preclinical studies showed that the surviving fraction of $MoS_2$-treated mice can appreciably increase to up to 79 % when they were exposed to high-energy ionizing radiation. Furthermore, $MoS_2$ dots can contribute in cleaning up the accumulated free radicals within the body, repairing DNA damages and recovering all vital chemical and biochemical indicators, suggesting their unique role as free radical scavengers. $MoS_2$ dots showed rapid and efficient urinary excretion with more than 80 % injected dose (I.D.) eliminated from the body after 24 hours due to their ultrasmall hydrodynamic size and did not cause any noticeable toxic responses up to 30 days.

**Introduction**

High-energy ionizing radiations (X-rays and gamma rays) are widely used in industry and medicine, but these radiations also cause significant health hazards such as cancer and other related diseases.[1-6] In clinical medicine, more than 50% cancer patients need to receive radiation therapies, but high-energy radiations during the treatments not only kill cancer cells but also cause inevitable damages to normal tissues.[7-8] Besides, more and more nuclear power plants are proposed in the world due to the rising demands in energy, imposing potential radiation risks on public health.[9-10] Under exposures to ionizing radiations, lots of free radicals including reactive oxygen species (ROS) are formed through ionizing reactions such as the photoelectric, Compton scattering and Auger effects.[11-12] These free radicals are able to react with DNA and RNA inside the body, cause structural and functional changes and affect biological processes.[13-15] As a result, it induces cell apoptosis and further triggers cancer or even death. The use of radioprotectants provides a feasible solution to shield heath tissues from high-energy radiations.[16] Amifostine (Ethyol®) is an extensively used prescription radioprotectant in radiation medicine by scavenging oxygen-derived free radicals, but its blood elimination half-life is only 1 minute, limiting its scavenging activities against ROS.[16] Other materials have shown abilities of radiation protection to an extent, but they cannot afford effective excretions which could result in potential hepatic and splenic toxicities.[14-15] As a matter of fact, US Food and Drug Administration (FDA) have required all injected agents for *in vivo* uses to be completely cleared from the body.[17] Therefore, it is highly desirable to explore an ideal radioprotectant with capabilities of highly efficient removals of ROS, renal clearance and low toxicities, for clinical translation as an adjuvant in radiotherapies.

To address these critical challenges, we utilized a simple approach towards the synthesis of ultrasmall cysteine-protected $MoS_2$ dots as radioprotectants. Surface protection with zwitterionic cysteine offers several indispensable merits to the MoS$_2$ dots. Firstly it avoids the significant increase in hydrodynamic sizes as compared to other surface ligands and thus allows effective elimination from renal clearance.[18-19] Secondly, the aqueous dispersibility and stability are significantly enhanced by the surface modification, maintaining the ultrasmall size *in vivo* and refraining from aggregation. Last, non-specific adsorption of serum proteins, especially opsonin, is largely prohibited, leading to postponed removal from the body and a relatively longer circulation time in blood to achieve desired radiation protection. The as-prepared MoS$_2$ dots exhibited extraordinary electrocatalytic activities for hydrogen peroxide and oxygen reduction reactions (ORRs), leading to lots of free electron transfers. Its endogenous catalytic properties provide a promising and effective pathway for scavenging free radicals *in vivo* via rapid reactions with oxygen radical superoxide ($O^{2-}$) and non-radical oxidant hydrogen peroxide ($H_2O_2$) in blood. Owing to their strong catalytic activities, MoS$_2$ dots improved the surviving fraction of mice exposed to high radiation doses of 662 keV gamma ray. Furthermore, the ultrasmall MoS$_2$ dots can indeed eliminate the ROS through reduction reactions in major organs. As a result, superoxide dismutase (SOD), as an important indicator for antioxidant defense in almost all living cells under exposure to oxygen, nearly recovered back to normal levels, suggesting the repair of radiation-induced damages. The ultrasmall cysteine-protected MoS$_2$ dots can be rapidly excreted via kidney and did not cause any toxicological responses in 30 days post injection.

**Results and Discussion**

Ultrasmall MoS$_2$ dots were synthesized and purified by a combinational approach of ultrasonication and gradient centrifugation according to our published procedures.[20-22] However, it is difficult to disperse bare MoS$_2$ dots as a stable colloid solution in water, preventing their further applications in biological systems. A cysteine protection layer with a molecular weight of

121 Da was introduced which not only maintains the ultrasmall hydrodynamic size to meet the cut-off size of <5.5 nm for renal clearance,[17] but also endows excellent biocompatibility in biomedical applications. Cysteine-protected $MoS_2$ dots were obtained and stabilized through the chemical bonding and van der Waals interactions between $MoS_2$ and cysteine,[23] and were further purified and re-dispersed in water (**Figure 1a**). Obvious Tyndall effects were observed, indicative of the existence of small particles in solution. **Figure 1b** exhibited a representative image of transmission electron microscopy (TEM) which showed cysteine-protected $MoS_2$ dots have a core size around 2 nm by analyzing 121 dots from TEM images (**Figure S1b**) and a hydrodynamic size of 3.1 nm (**Figure S1c**) by dynamic light scattering (DLS). X-ray photoelectron spectroscopic (XPS) investigation of cysteine-protected $MoS_2$ manifested Mo and S elements from $MoS_2$ and N element from cysteine, suggesting the formation of cysteine-$MoS_2$ composite (**Figure 1c**). Additionally, the S $2p_{3/2}$ binding energy located at 164.0 eV suggested the formation of disulfide bonds which are ascribed to the conjugation of cysteine and $MoS_2$ dots (**Figure 1d**).[24] On the other hand, no signals from disulfide bonds can be found on the corresponding counterparts, cysteine and unprotected $MoS_2$ dots (**Figure S2**). Optical absorption was investigated by UV-vis spectroscopy (**Figure 1e**). Cysteine alone did not show any absorption features, while the $MoS_2$ suspension and cysteine-protected $MoS_2$ dots displayed multiple absorption peaks at 393, 470, 607 and 670 nm, attributed to deep energy levels of electronic transitions and interband excitonic transitions.[25-26] Notably, the results are similar to the absorption bands observed for $MoS_2$ particles prepared by other methods.[25-26]

We evaluated the *in vitro* catalytic activities of cysteine-protected $MoS_2$ dots towards reduction of $H_2O_2$ in $N_2$-saturated 0.01 M phosphate-buffered salin (PBS), pH = 7.4, using a cysteine-$MoS_2$ modified glassy carbon electrode (GCE). **Figure 1f** showed the cyclic voltammetric (CV) curves of the ultrasmall cysteine-protected $MoS_2$ dots in the presence and absence of 5.00 mM $H_2O_2$ at the scan rate of 50 mV·s$^{-1}$. Negligible reduction current was observed in the absence of $H_2O_2$, whereas in the presence of $H_2O_2$, the reduction current increased sharply when the scan potential shifted to more negative than -0.4 V, indicating extraordinary catalytic activities toward reduction of $H_2O_2$ compared to voltammetric responses of the bare GCE without modification of cysteine-protected $MoS_2$ dots (**Figure S3a**). To determine the role of cysteine-protected $MoS_2$ dots in oxygen reduction reaction (ORR), CV responses of cysteine-protected $MoS_2$ dots were investigated in $O_2$-saturated 0.01 M PBS solution at the scan rate of 50 mV·s$^{-1}$ (**Figure 1g**). Cysteine-protected $MoS_2$ dots showed electrocatalytic activity for reduction of $O_2$ with a more positive onset potential and larger current density compared to those of the unmodified GCE (**Figure S3b**). The electrochemical measurements of cysteine-protected $MoS_2$ dots evidenced their strong *in vitro* catalytic activities in $H_2O_2$ and oxygen reduction reactions, serving as the original inspiration for us to investigate the biological responses in radiation-injured mice.

We firstly examined the *in vitro* toxicities of cysteine-protected $MoS_2$ dots in 3T3/A31 cells using neutral red assay and MTT method which showed extremely low cytotoxicities with doses up to 145 μg/ml (**Figure S4**). Viabilities of cells treated with cysteine-protected $MoS_2$ dots in different doses under exposures to gamma ray were significantly increased compared with those without treatments (**Figure 2a**), suggesting the powerful *in vitro* protective behaviors of cysteine-protected $MoS_2$ against radiation. The DNA damages of cysteine-protected $MoS_2$ dots with different concentration up to 145 μg/ml also investigated by single-cell sol-gel electrophoresis, and no significant difference on tail moments was found, suggesting negligible DNA damages from $MoS_2$ dots (**Figure S5**). DNA damages arising from radiation were further estimated (**Figure S6**). No obvious cell tail moments were found in the control, but cell tails implying DNA damages were easily identifiable with exposures to high-energy gamma ray. In contrast, the cell tails from cells treated with cysteine-protected MoS$_2$ dots showed distinct recoveries, presenting effective DNA repairs. Quantitative investigations of tail moments showed that cells treated with cysteine-protected MoS$_2$ dots recovered to an appreciable degree from exposures of gamma ray, as compared to cells treated only with radiation (**Figure 2b**).

*In vivo* protection from radiation using cysteine-protected MoS$_2$ dots were investigated with C57BL/6 mice. 200 μL cysteine-protected MoS$_2$ dots at different concentrations were intraperitoneally injected into mice. The mice were then exposed to high energy gamma ray ($^{137}$Cs, 3600 Ci) at the dose of 7.5 Gy. The surviving fractions of radiation-injured mice with various injection concentrations are shown in **Figure 2c**. The surviving fraction of mice exposed to high-energy radiations without injection of cysteine-protected MoS$_2$ dots is 0 after two weeks. This result demonstrated radiation induced acute damages and death which are consistent with observations in previous published work.[6] With injection of cysteine-protected MoS$_2$ dots, the surviving fractions of mice were 7.1 and 42.9 and 78.6 % at the doses of 10, 20 and 50 mg/kg. Meanwhile, the surviving fraction of mice treated with cysteine-protected MoS$_2$ dots (200 μL, 5 mg/ml) in the absence of radiation is 100 %, indicating low toxicities at this concentration. In fact, cysteine protection layers were observed to show outstanding biocompatibility, in good agreement with previous results.[27-28] Besides, we also investigated the *in vivo* radiation protection effects of Pt-, Co- and Ni-doped MoS$_2$ dots as well as cysteine, but none of these metal-doped MoS$_2$ dots showed comparable survival rates to that of cysteine-protected MoS$_2$ dots (**Figure S7**), which may be due to the large sizes and different surface chemistries. Noticeably, a low survival rate was observed for cysteine-treated mice, clearly providing strong evidences that the radiation protection effect against gamma ray is from MoS$_2$ instead of cysteine. Next, DNA damages of irradiated mice treated with or without cysteine-protected MoS$_2$ were assessed (**Figure 2d**). Total DNA from bone marrow cells and bone marrow nucleated cells (BMNC), two dominant indicators of ionizing radiation, were collected from control mice, irradiated mice and irradiated mice treated with cysteine-protected MoS$_2$. DNA from healthy mice showed an optical density (OD) of 0.42 after 1 day, but it drastically decreased to 0.22 only after 1 day post exposure to radiation, indicating severe DNA damages induced by radiation. In contrast, OD of total DNA for mice treated with cysteine-protected MoS$_2$ dots decreased to a higher value of 0.29 after 1 day. After 7 days, OD of mice treated with cysteine-protected MoS$_2$ dots rose to 0.38, almost recovered to the healthy value, compared to that of 0.25 from mice with only radiation. These results illustrate that cysteine-protected MoS$_2$ dots can effectively decrease DNA damages from high-energy radiations. The number of BMNC is presented as in **Figure S8**. Similar to the results of DNA damages, the BMNC number decreased from 6×10$^6$ cells/mL in healthy mice to 4.3×10$^6$ cells/mL in irradiated mice after 1 day, while it was recovered to 4.6×10$^6$ cells/mL in irradiated mice treated with cysteine-protected MoS$_2$ dots. On the 7$^{th}$ day, the BMNC number in mice treated with MoS$_2$ dots was stably maintained at 4.6×10$^6$/mL as to a significantly decreased level of 1.3×10$^6$/mL in mice only treated with radiation. The results clearly presented that cysteine-protected MoS$_2$ dots are critical in recovering the number of BMNC.

To address long-term damages caused by radiation, blood chemistry panels of irradiated mice under 7 Gy gamma ray were studied (**Figure S9**). In the beginning (1 day), it is clear that white blood cells (WBC), red blood cells (RBC) and platelets (PLT) sharply decreased after exposure to gamma ray indicating strong radiation-induced inflammatory responses, independent of treatments of cysteine-protected MoS$_2$ dots. At the middle time point (7 days), WBC and PLT were still in low levels without distinct recoveries to healthy levels, while all other indicators started to display certain recoveries after treatments with cysteine-protected MoS$_2$ dots. After 30 days, a large number of the indicators of mice treated with irradiation showed significant decreases, but those receiving treatments of cysteine-protected MoS$_2$ dots have appreciably recovered to normal levels. Besides, radiation triggered significant changes in levels of alanine aminotransferase (ALT) and serum creatinine (CERA) 1 and 7 days post injection (**Figure S10**). On the other hand, all these indicators for $MoS_2$-dot-treated mice were thoroughly restored to normal 30 days post injection, while irradiated mice without $MoS_2$ treatments presented significant variations in the levels of ALT, albumin (ALB), blood urine nitrogen (BUN) and globulin (GLOB). Under exposure to radiations, the acute radiations not only give rise to severe DNA breaks and considerable decrease in the BMNC numbers, but also instantaneously elicit inflammatory responses.[29] Radiation-induced damages were the most severe after 7 days and lots of irradiated mice even began to die. After that, the mice started to recover gradually up to the 30th day. Undoubtedly, radiation always induces certain irreparable damages of DNA and BMNC and thus strongly affects the panels of blood chemistry and biochemistry, implying infections and inflammations. However, the cysteine-protected $MoS_2$ dots can free the mice from radiation-related DNA and BMNC damages, relieve corresponding infections and thus increase the overall surviving fractions of mice.

To further reveal the mechanism of the underlying radiation protection, superoxide dismutase (SOD) and 3,4-Methylenedioxyamphetamine (MDA) were employed to provide some insights. SOD are enzymes that alternately catalyze dismutation (or partitioning) of the superoxide ($O_2^-$) radicals into either ordinary molecular oxygen ($O_2$) or hydrogen peroxide ($H_2O_2$). In contrast, MDA is a harmful product generated by mice. The contents of SOD and MDA indicate levels of damages caused by ionizing radiation. **Figure 3a** and **b** showed SOD levels in lung and liver. When mice were exposed to gamma ray, the SOD of lung from all mice, independent of treatments of $MoS_2$ dots, notably decreased after 1 day. However, 7 days post radiation, the SOD from the mice treated with cysteine-protected $MoS_2$ dots exhibited powerful recoveries compared to mice without treatments, clearly elucidating the increase in the ability of

ROS clearance. **Figure 3c** and **d** examined the MDA levels of lung and liver from irradiated mice with or without treatments of cysteine-protected $MoS_2$ dots. The MDA level in the lung of normal mice was 8.1 nmol/mL, but it sharply increased to 22.3 nmol/mL after 1 day (**Figure 3c**). In the meantime, the treatment with cysteine-protected $MoS_2$ dots could partially shield the mice from radiation damages with much lower MDA levels of 15.1 nmol/mL after 1 day. After 7 days, the MDA levels decreased to 10 nmol/mL for mice with treatments of cysteine-protected $MoS_2$ dots compared to 12.8 nmol/mL for mice only with radiation. Similarly, **Figure 3d** indicated the MDA levels in the liver sharply increased from 3.6 nmol/mL to 13.2 and 12.1 nmol/mL for treated and untreated mice, respectively. After 7 days, the mice treated with cysteine-protected $MoS_2$ dots recovered back to 4 nmol/mL, but untreated mice was still 5.5 nmol/mL, significant higher than the threshold value of normal mice without irradiations. In healthy mice, SOD is maintained at normal levels,[30-31] but high-energy irradiation is well known to cause generation of excessive amount of ROS and consumption of large quantities of SOD. Consequently, the SOD levels from liver and lung decrease enormously. The cysteine-protected $MoS_2$ dots can help to get rid of the undesirable hazardous ROS *in vivo* and rescue the body from consuming lots of SOD to achieve the removal of ROS and accordingly, SOD can be recovered and reserved at the normal level. Similarly, MDA is normally present in low levels in the body. When mice suffered from the exposure to radiation, a large quantity of MDA is produced due to the generation of ROS in excess. The cysteine-protected $MoS_2$ dots can decrease MDA levels of mice via eliminating free radicals. As such, as the cysteine-protected $MoS_2$ dots with reducing potencies circulate into blood, they reduce the radiation-induced ROS, functioning as free radical scavengers.

For further medical applications, pharmacokinetics and biosafety are the two most important characteristics to be taken into consideration. We investigated the pharmacokinetics, urine excretion and *in vivo* toxicities of cysteine-protected MoS$_2$ dots at the injection dose of 50 mg/kg. The C57/BL6 mice were injected at the concentration of 5 mg/ml with blood and urine collected for pharmacokinetic and renal-excretion studies, as measured by inductively coupled plasma mass spectrometry (ICP-MS). **Figure 4a** shows that the cysteine-protected MoS$_2$ dots has a half time of 2.1 hours in blood. Furthermore, nearly 80% of cysteine-protected MoS$_2$ dots can be rapidly excreted through the urine route after 24 hours post injection due to their ultrasmall hydrodynamic size (**Figure 4b and Figure S11**). Biodistribution of cysteine-protected MoS$_2$ dots was analyzed after 24 hours post injection using ICP-MS. Bladder and kidney showed the highest distribution, ascribed to renal clearance, whereas spleen, liver, lung and heart had relatively low uptakes (**Figure 4c**). With increasing post injection time of up to 30 days, the uptakes of MoS$_2$ dots from all organs sharply decreased as compared to 1 day and MoS$_2$ dots have almost been completely eliminated from the body. Traditional nanoparticles with larger sizes cannot be fast excreted, inducing high uptakes in liver and spleen.[32] For example, carbon nanotubes with PEG coatings can only be excreted for a total of 60% after 90 days post injection,[33-34] and lots of nanotubes accumulated in liver and spleen will induce potential liver toxicities. Noticeably, radiation-protective effects of CeO$_2$ and Ag nanoparticles have been previously reported, but all those nanomaterials are not clearable by the glomerular filtration of kidney, owing to their large sizes as well as active surface chemistries.[15, 35] The ultrasmall cysteine-protected MoS$_2$ dots shown here, nevertheless, provide the unique feature of highly efficient renal clearance, resembling small molecules.

We next evaluated the *in vivo* toxicities of cysteine-protected MoS$_2$ dots in terms of body weight, immune responses, hematology and biochemistry panels at the time points of 1, 7 and 30 days. During the 30-day period, treatments with cysteine-protected MoS$_2$ dots did not induce any obvious adverse effects on the growth of mice and no meaningful statistical differences were observed in the body weight and the thymus index between the cysteine-protected MoS$_2$ dots-treated mice and control mice. We examined standard hematological biomarkers including WBC, RBC, hematocrit (HCT), mean corpuscular volume (MCV), hemoglobin (HGB), PLT, mean corpuscular hemoglobin (MCH) and mean corpuscular hemoglobin concentration (MCHC). Hematological results for the cysteine-protected MoS$_2$ dots are presented in **Figure 4d and Figure S12a**. For the mice treated with cysteine-protected MoS$_2$ dots, two typical indicators (WBC, RBC) did not show any meaningful differences from those of untreated mice. These results clearly illustrate that the cysteine-protected MoS$_2$ dots do not induce significant infections and inflammations in mice and are relatively safe in pre-clinical settings.

In addition, we performed the standard biochemistry examination on the mice treated with cysteine-protected MoS$_2$ dots at different time points of 1, 7 and 30 days (**Figure 4e and Figure S12b**). The biochemical parameters including ALT, AST, total protein (TP), ALB, blood urea nitrogen, CREA, GLOB, and total bilirubin (TBIL) were investigated. We emphasize ALT, AST, and CREA, because they are closely related to the functions of liver and kidney of mice. ALT showed increase after 7 days, but recovered to normal after 30 days treatments. AST and CREA did not have statistical differences in the treated mice. In fact, MoS$_2$ nanosheets and other two-dimensional nanomaterials have shown potential applications in imaging and photothermal therapies, but the hydrodynamic sizes of these particles are too much larger than the 5.5 nm cut-off value of kidney, which will induce potential liver toxicities.[36-43] For example, Au nanoparticles coated by PEG and BSA caused acute damages even after 30 days at a relatively low injection dose of 5 mg/kg,[44-45] In comparison, the sub-5 nm MoS$_2$ dots we prepared with efficient renal clearance will escape from the uptake of the reticuloendothelial system (ROS), minimizing the toxic effects. The cysteine-protected MoS$_2$ dots showed extremely low liver toxicities even at a 10-fold high dose of 50 mg/kg. Finally, the pathological changes of organs were demonstrated by immunohistochemistry at different time points. Liver, spleen, kidney and other organs were collected and sliced for Haematoxylin and Eosin (H&E) staining (**Figure 4f and Figure S13**). No apparent damages were observed in all organs, especially the spleen and kidney during the entire period.

Radiation protection is utmost important in both healthcare and environment. Fukushima nuclear radiation has caused detrimental public crisis and irreversible radiation damages in large areas.[46] Thus, the development of radioprotectants with low toxicities or even without toxicities is promising for potential applications in this area. Cysteine-protected $MoS_2$ dots not only endow renal clearance, but also offer strong catalytic properties that allow to react with excessive radiation-induced ROS in the blood of mice. However, we have conceived that it still has large room for the development of new materials for protection from ionizing radiation. In the future, it is valuable to explore new small-molecular organic materials with high activities in free radical scavenging. It is necessary to develop effective radioprotectants with sub-5 nm hydrodynamic sizes as well as low toxicities.[47-54] Besides, it is of great importance to design materials with well-defined functional surfaces for radiation protection. It is also interesting to combine features of energy transfers and energy-dependent responses under exposure to high-energy radiation. Detailed biological and bio-catalytic mechanisms of radiation protection are still necessary to be further investigated.[55-57]

**Conclusion**

In summary, we developed ultrasmall cysteine-protected $MoS_2$ dots radioprotectants, with high efficiencies in renal clearance and radiation protection against gamma ray. This type of novel nanomaterials behave like organic molecules with low levels of retentions in RES organs such as liver and spleen, avoiding associated toxicities. The cysteine-protected $MoS_2$ dots also manifested high catalytic activities against $H_2O_2$ and $O_2$, leading to potential removal of ROS *in vivo*. Both *in vitro* and *in vivo* experiments showed that the surviving fractions of mice can be significantly increased with exposure to high-energy gamma ray. Furthermore, treatments of cysteine-protected $MoS_2$ dots can considerably decrease DNA breaks and BMNC damages and can also recover radiation-induced damages on WBC, PLT and other vital chemical and biological indicators. The cysteine-protected $MoS_2$ dots behave as free radical scavengers and induce increase in SOD and decrease in MDA. Finally, we found cysteine-protected $MoS_2$ dots achieved around 80% urine excretion via bladder in 24 hours without any observable toxic effects even at a high injection dose of 50 mg/kg.

**Experimental Section**

**Preparation of Ultrasmall Cysteine-protected MoS$_2$ Dots.** Firstly $MoS_2$ dots were synthesized by a series of ultrasonication and centrifugation steps according to our previous work[20]. $MoS_2$ powder (99%, <2 μm in size, Aldrich) was dispersed in N,N-dimethylformamide (DMF) (99.9%, Aldrich) at the concentration of 1 mg mL$^{-1}$ and then subject to ultrasonication for 8 h using a SB-2200 sonifier (Shanghai Branson, China) to form a black suspension. The resulting suspension was centrifuged at 6000 rpm for 30 min using a high-speed centrifuge (Changsha Pingfan Instrument and Meter Co., Ltd., China), with the light yellow supernatant collected. It was then centrifuged again at 12000 rpm for 30 min and the precipitated pellets were washed with deionized water several times and redispersed in water to form an aqueous suspension of $MoS_2$ dots as the unprotected control. As for cysteine-protected $MoS_2$ dots, the precipitates were re-dispersed into DMF of the same volume as before centrifugation and L-cysteine (99%, Beijing Chemical Reagent Co., Ltd.) was added to achieve a final concentration of 0.1 mg mL$^{-1}$. The mixture was mixed uniformly and kept still for 24 hours of aging at room temperature. Particles in larger sizes were removed by centrifugation at 6000 rpm for 15 min and the remaining supernatant was centrifuged again at 12000 rpm for 30 min with the precipitates collected. The product was washed extensively with deionized water and centrifuged at 12000 rpm to remove free cysteine and residual DMF. The solution of cysteine-protected $MoS_2$ dots was prepared by redispersing the precipitates into certain volumes of water.

**Preparation of Pt-, Co- or Ni-doped $MoS_2$.** The Pt, Co and Ni doped $MoS_2$ (Pt-$MoS_2$, Co-$MoS_2$ and Ni-$MoS_2$) were synthesized according to our previous work.[58] For Pt-$MoS_2$, 900 mg $(NH_4)_6Mo_7O_{24}\cdot 4H_2O$ and 0.442 mL 0.19 mol $L^{-1}$ $H_2PtCl_6$ (aq.) were dissolved in 20 mL deionized water to form a homogeneous solution, which were then transferred into a 40 mL stainless steel autoclave with 10 mL $CS_2$ under Ar and maintained at 400 °C for 4 h. The saturated NaOH (aq.) were used to treat the obtained product at 60 °C for 3 h, followed by washing several times with water and absolute ethanol and drying at 100 °C. The Co–$MoS_2$ and Ni–$MoS_2$ were synthesized by using 900 mg $(NH_4)_6Mo_7O_{24}\cdot 4H_2O$ and 0.085 g $Co(NO_3)_2\cdot 6H_2O$ or 0.084 g $Ni(NO_3)_2\cdot 6H_2O$ dissolved in 20 mL deionized water and 10mL $CS_2$, following the same process as for Pt–$MoS_2$. The doping contents of Pt, Co and Ni in $MoS_2$ were all 1.7 wt.% measured by inductively coupled plasma optical emission spectroscopy (ICP-OES).

**Modification of GCE with Ultrasmall Cysteine-protected $MoS_2$ Dots.** Cysteine-protected $MoS_2$ dots were redispersed in water at the concentration of 1 mg $mL^{-1}$. Prior to modification, GCE was polished with 0.3 μm alumina slurries followed by 0.05 μm alumina slurries (Buehler). The polished GCE was sonicated three times, each for 3 min in deionized water to remove any residual polishing reagents. The solution of $MoS_2$ was drop casted on GCE (diameter = 3 mm) and dried in air with a loading of 0.2 mg $cm^{-2}$.

**Characterization.** All electrochemical measurements were carried out on a CHI 660D electrochemical workstation (Chenhua, China) at room temperature. The cysteine- $MoS_2$ dot-modified GCE, a Pt electrode and a saturated calomel electrode (SCE) were used as the working, counter and reference electrodes respectively for all electrochemical measurements. Transmission electron microscopy (TEM) images were acquired on a JEM-2100F electron microscope (JEOL, Japan). X-ray photoelectron spectra (XPS) were collected on an Axis Ultra spectrometer (Kratos Analytical Ltd., Japan) and the binding energy was calibrated by the C 1s peak at 284.8 eV. UV-vis absorption spectra were recorded with a U-4100 UV-vis-NIR spectrophotometer (Hitachi, Japan). The hydrodynamic diameter of the cysteine-protected $MoS_2$ nanodots was determined by dynamic light scattering (DLS) with a NanoZS Zetasizer (Malvern). The sample solutions were prepared by dissolving the cysteine-protected $MoS_2$ nanodots in deionized ultrafiltered (DIUF) water at a concentration of 0.01 mg/mL. DLS data were acquired in the phase analysis light scattering mode at 25 °C and poly-dispersion index is always <0.7 for every measurement.

*In Vitro* **Cytotoxicity and Radiation Protection**. BALB/3T3 clone A31 mouse fibroblasts (A31) were employed for cell viability experiments. A31 cells were dispensed in two 96-well plates with $7 \times 10^3$ cells per well and incubated under 37℃, 5% $CO_2$ and humidified atmosphere. Different concentrations of $MoS_2$ from 0.5-135 μg/ml were introduced into the DMEM culture media. After incubation for 24 and 48 hours, cytotoxicities were analyzed with Neutral Red (NR) staining. Meanwhile, to further confirm the accuracy of the cell survival rates, cysteine-protected $MoS_2$ nanodots were also assessed using MTT Cell Proliferation and Cytotoxicity Assay Kit. Cysteine-protected $MoS_2$ dots (final concentrations of 0.5-135 μg/ml) were introduced into the DMEM culture media. After 24 or 48 h of treatments, 10 μL of MTT reagent was added to each well and incubated for 4 h, and then the media were replaced with 150 μL DMSO to dissolve formazan crystals. The optical absorption in 490 nm was measured with a single tube luminometer (TD 20/20, Turner Biosystems Inc., Sunnyvale, CA, USA) with the intrinsic optical absorption of cysteine-protected $MoS_2$ nanodots subtracted. The cell survival rates were calculated based on the recorded optical absorption.

$4 \times 10^3$ A31 cells were seeded into five 96-well plates, supplemented with high-glucose DMEM containing 10% FBS. After 10 hours of incubation, the cells were treated with 0.2 µg and 2 µg $MoS_2$ dots for 1 hour which were dissolved in 100 ul complete culture media. The five plates were then exposed to 0, 2, 4, 8 and 10 Gy radiations respectively. After 48 hours, NR was used to evaluate cell viabilities. When cells had sufficiently taken up NR in about~3 hours, the media containing NR was discarded and each well was gently washed with 1X PBS, followed by addition of 150 µl desorb solution to extract NR from living cells. All the plates were homogenized on a microtiter plate shaker for 5 min and the optical absorbance was recorded at 540 nm by Infinite F200 multimode plate reader.

*In Vitro* **DNA Damages**. The in vitro protocols were approved by the Institutional Review Board of Institute of Radiation Medicine at the Chinese Academy of Medical Sciences (CAMS). A modified protocol of alkaline COMET assay was used to evaluate DNA damages. Agarose (0.8%, 500 µl) was paved homogeneously on glass microscopic slides. After solidification, $4.8 \times 10^4$ cells in 30 µL 1X PBS incubated with $MoS_2$ dots (0.5-135 µg/ml) were mixed with 70 µL low-melting-point agarose (0.6%) and 20 µl of this mixture was spread over the slide completely. The solidified slides were placed into freshly-prepared ice-cold lysis buffer (2.5 mol/L NaCl, 100 mmol/L $Na_2EDTA$, 10 mmol/L Tris-HCl, 10% DMSO, 1% Triton X-100) for 2 hours, followed by immersing them in a horizontal gel electrophoresis unit which was filled with chilled electrophoresis buffer (1 mmol/L $Na_2EDTA$, 300 mmol/L NaOH, pH 7.4 for 30 min. Electrophoresis was conducted at 30 V for 20 min. The slides were neutralized with ethanol after electrophoresis and stained with ethidium bromide (2 µg/mL). DNA damages were analyzed by Comet Assay Software Project (CASP) for tail moments. Typically, 100 cells were analyzed for DNA tail moment.

***In Vivo* Radiation Protection**. All animal-related protocols were reviewed and approved by the Institutional Animal Care and Use Committee (IACUC). All animals were purchased, maintained and handled under protocols approved by the Institute of Radiation Medicine at the Chinese Academy of Medical Sciences (CAMS). 200 μL saline was intraperitoneally injected into mice as the control group. As for the treatment group, 1, 2 or 5 mg/mL cysteine-protected $MoS_2$ dots were used for animal experiments by intraperitoneal injection with doses of 10, 20 and 50 mg/kg in mice. Subsequently, the mice were radiated under 7.5 Gy gamma rays (~8 min) and $^{137}Cs$ with an activity of 3600 Ci and a photon energy of 662 KeV were used. 70 mice in total were assigned into the following 5 groups (14 mice in each group): control, only treated with radiation, treated with both radiation and cysteine-protected $MoS_2$ dots at the doses of 10, 20 and 50 mg/kg. We then monitored all the mice for up to 30 days and the surviving fractions of mice were obtained for each group.

**Analysis of Total DNA and Bone Marrow Nucleated Cells:** 32 male C57BL/6 mice were assigned into the following 4 groups (8 mice each group): control of 1 day, treatment group of 1 day, control of 8 days and treatment group of 8 days. For the control groups, mice were administered with 200 μL distilled water and the treatment groups were intraperitoneally injected with 200 μl solution of cysteine-protected $MoS_2$ dots at the dose of 50 mg/kg. The entire body of each mouse was radiated under gamma ray at the dose of 6.5 Gy. After 1 and 8 days of treatments, mice were sacrificed. Bilateral femurs of each mouse were excised from the body and connective tissues were removed completely. To estimate the total DNA levels in bone marrows, bone marrow cells were flushed from the femurs into calcium chloride solution (10ml, 5 mM) with a 24-gauge needle and as such, single-cell suspensions were made. The suspensions were placed at 4℃ for 30 minutes and centrifuged at 2500 rpm/min for 15 minutes with supernatants discarded. The pellets were mixed with perchlorate (5 mL 0.2 M) and stored in water bath being heated for

15 min at 90℃. After being cooled down to room temperature, the mixtures were purified through filter paper (pore size =0.2 μm) and the filtrates were measured for UV-vis absorption at 268 nm (Shimadzu, UV-1750). Similarly, bone marrow nucleated cells were flushed into 1 mL 1X PBS, filtered through nylon meshes to remove fragments of bones and tissues for cytometric analysis (Mindray BC-2800 Vet).

**SOD and MDA Analysis**: All the organs for SOD and MDA analysis were from identically treated mice as for total DNA and BMNC analysis. After 1 and 8 days of treatments, mice were sacrificed with livers and lungs collected for analysis of SOD and MDA levels using Total Superoxide Dismutase assay kits and Malondialdehyde assay kits (Nanjing Jiancheng Bioengineering Institute). All the organs were immersed in saline solution and homogenized with a tissue homogenate machine (IKA, T18 basic) to make 10% tissue homogenate. The organ homogenates were left on ice for 1 hour and then centrifuged at 3500 r/min for 10 minutes. The resulting solutions were further diluted into 1 % and 0.25 % homogenate respectively. For total SOD level analysis: phosphate buffer (1 ml, 7.5 mM) was added to 5 ml EP tubes and then mixed with liver and lung homogenate (50 μl, 0.25 %) or 50 μl distilled water as control. Hydroxylamine hydrochloride (0.1 M, 100 μl), xanthine solution (75 mM, 100 μl) and xanthineoxidase solution (0.03 U/L, 100 μl) were added into the above EP tubes in order, fully mixed and incubated at 37℃ for 40 minutes. After that, 2ml nitrite developer ($C_{10}H_9N$: $C_6H_7NO_3S$: $CH_3COOH$=3:3:2) was added into the system and analyzed by UV-vis spectrophotometer (Shimadzu, UV-1750) at 550 nm. For MDA analysis, 10% tissue homogenates were employed. Equal volumes (100 μl) of organ sample, ethanol and tetrathoxypropane (10 nmol/ml) were added into individual 5 ml EP tubes, among which ethanol and tetrathoxypropane were negative and positive controls respectively. 100 ul tissue lysis solution was added into each 5ml EP tube and mixed completely. Trichloroacetic acid (10 %, 3 ml) and thiobarbituric acid (0.6 %, 1 ml) were added into the system orderly. Then the tubes were heated in water at 95℃ for 40 minutes. After being cooled down to room temperature, all the samples were centrifuged at 3500~4000 r/min for 10 minutes with the supernatants analyzed by UV-vis spectrophotometer at 532 nm (Shimadzu, UV-1750).

*In Vivo* **Toxicity**. Animals were purchased, maintained, and handled with protocols approved by the Institute of Radiation Medicine, Chinese Academy of Medical Sciences (IRM, CAMS). 48 male C57BL/6 mice at the age of 11 weeks were obtained from IRM laboratories, housed by 2 mice per cage with a 12 hours/12 hours light/dark cycle and were provided with food and water. Mice were randomly divided into 6 groups (8 mice in each group): control of 1 day, treatment group of 1 day, control of 7 days, treatment group of 7 days, control of 30 days, and treatment group of 30 days respectively. The solution of cysteine-protected $MoS_2$ dots (200 μL, 5 mg/mL) was used for this animal experiment through intraperitoneal injection. The assigned dose was 50 mg/kg in each mouse. Mice were weighed and assessed for behavioral changes every day post injection. After treatments with cysteine-protected $MoS_2$ dots for 1, 7 and 30 days, mice were sacrificed accordingly at each time point using isoflurane anesthetic and angiocatheter exsanguination with 1X PBS. Blood and organs were collected for biochemistry and pathological analyses. One mouse from each group was fixed with 10% formalin in PBS following exsanguination with PBS. During necropsy, liver, kidneys, spleen, heart, lung, testis, brain, bladder and thymus were collected and weighed.

**Hematology, Biochemistry and Pathology**: Using a standard saphenous technique, vein blood was collected in potassium EDTA collection tubes for hematological analysis. Standard hematological and biochemical examinations were performed. 1 mL of blood was collected from mice and separated into cellular and plasma fractions by centrifugation. Mice were sacrificed by isoflurane anesthetic and angio catheter exsanguinations. Major organs were harvested, fixed in 10% neutral buffered formalin, processed routinely into paraffin and stained by H&E, with pathology examined using a digital microscope.

**Statistical Analyses**: Paired Student's t-test was used for statistical analysis between the control group and the treatment group ($MoS_2$ alone or radiation+$MoS_2$). The Analysis of Variance (ANOVA) was used for statistical analysis among the control, radiation, and radiation+$MoS_2$ groups. Differences were considered statistically significant at a P value < 0.05.

ACKNOWLEDGMENT

This work was supported by the National Natural Science Foundation of China (Grant No.81471786 and 21475003 and 21275010), Natural Science Foundation of Tianjin (Grant No. 13JCQNJC13500).

**Author contributions.**

X.Z. and M.L, conceived and designed the experiments. J.W., J.Z., W.L., and J.C., X.S., J.D., performed the all experiments. J.W., X.S., and W.L., contributed to radiation protection. J.Z., and .J.D, contributed to synthesis. X.Z., M.L, J.Y, J.W., J.Z., W.L., J.C., X.S., J.D., D. D., and Y.S. analyzed the data and wrote the manuscript. All authors discussed the results and commented on the manuscript.

# Figures and Figure Captions

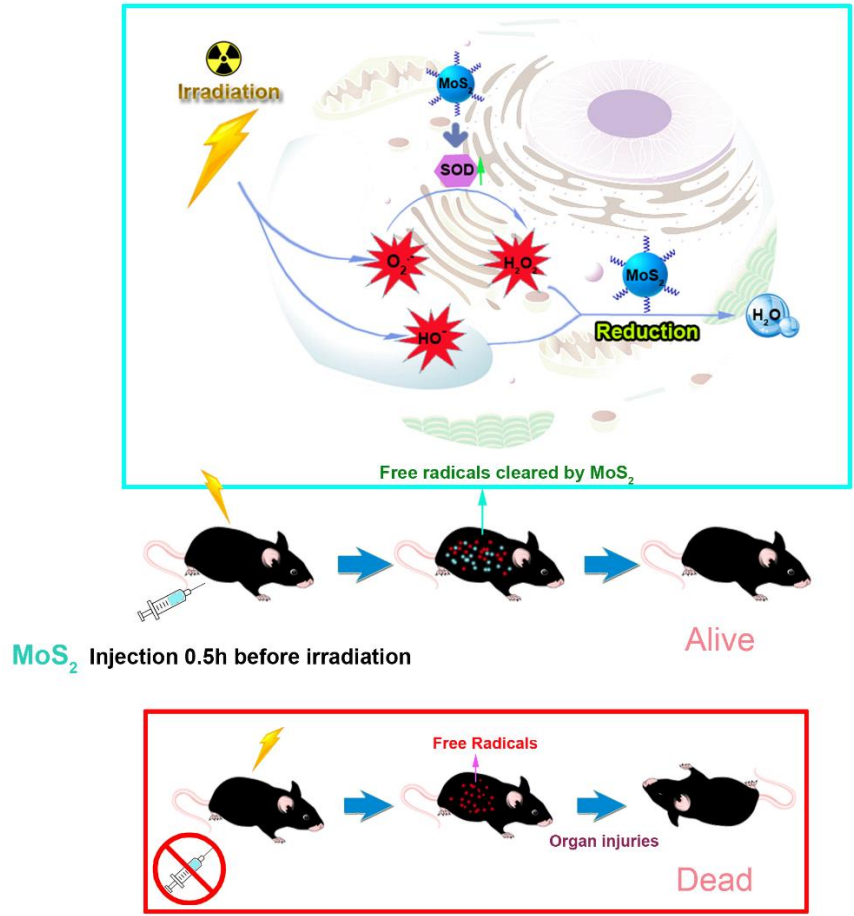

Schematic illustration of radiation protection with cysteine-protected MoS$_2$ dots

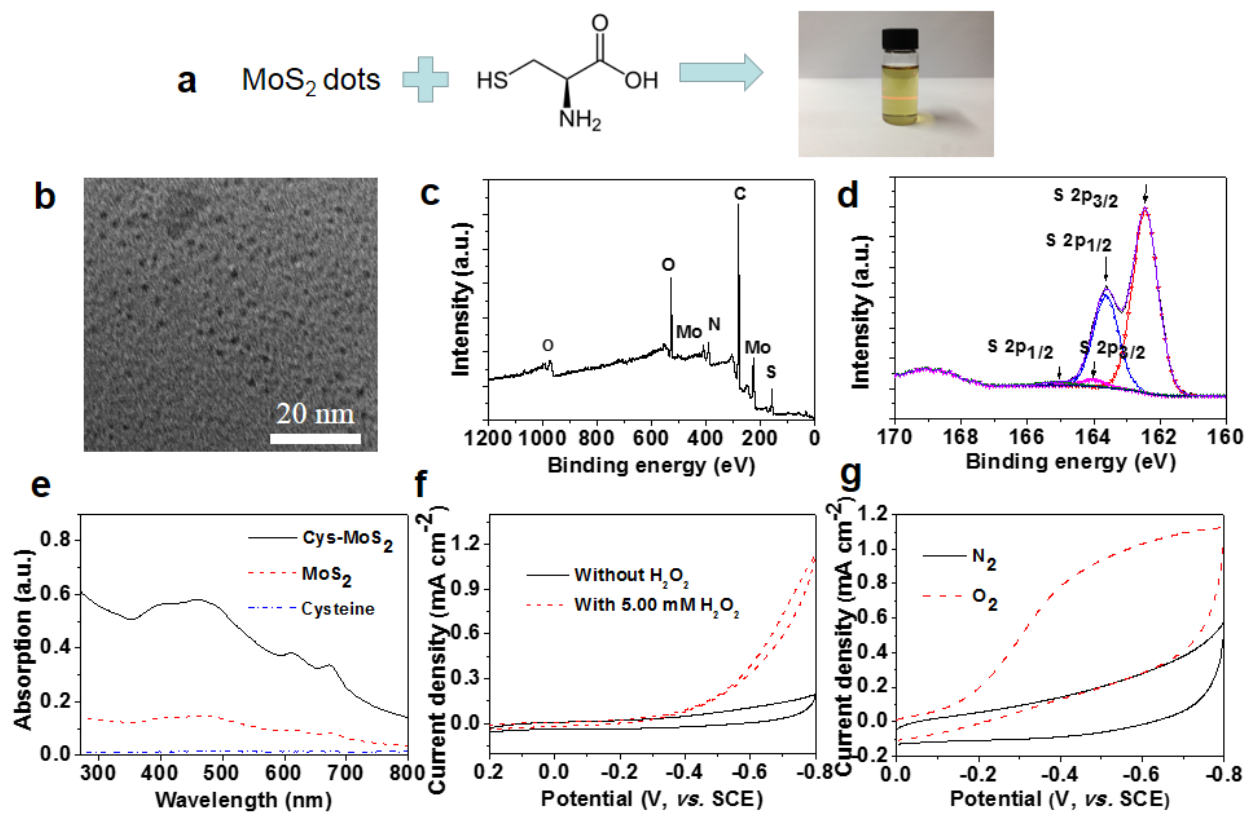

**Figure 1 Characterization and electrocatalytic properties of cysteine-protected MoS₂ dots.** (**a**) Schematic preparation of cysteine-protected MoS₂ dots. (**b**) TEM image of a population of cysteine-protected MoS₂ dots with a homogeneous distribution of around 2 nm. Scale bar, 20 nm. (**c**) Wide survey X-ray photoelectron spectrum and (**d**) S 2p spectrum of cysteine-protected MoS₂ dots. The presence of disulfide bonds indicates the bonding between cysteine and MoS₂. Arrows show corresponding electronic states in MoS₂ dots. (**e**) UV-vis spectra of aqueous solutions of cysteine, unprotected and cysteine-protected MoS₂ dots. (**f**) CVs of GCE modified with cysteine-protected MoS₂ dots in the presence (dotted) and absence (solid) of 5.00 mM $H_2O_2$ in $N_2$-saturated 0.01 M pH 7.4 PBS. Scan rates: 50 mV s$^{-1}$. (**g**) CVs of GCE modified with cysteine-protected MoS₂ dots in $N_2$- (solid) and $O_2$-saturated (dotted) 0.01 M pH 7.4 PBS. Scan rate: 50 mV s$^{-1}$.

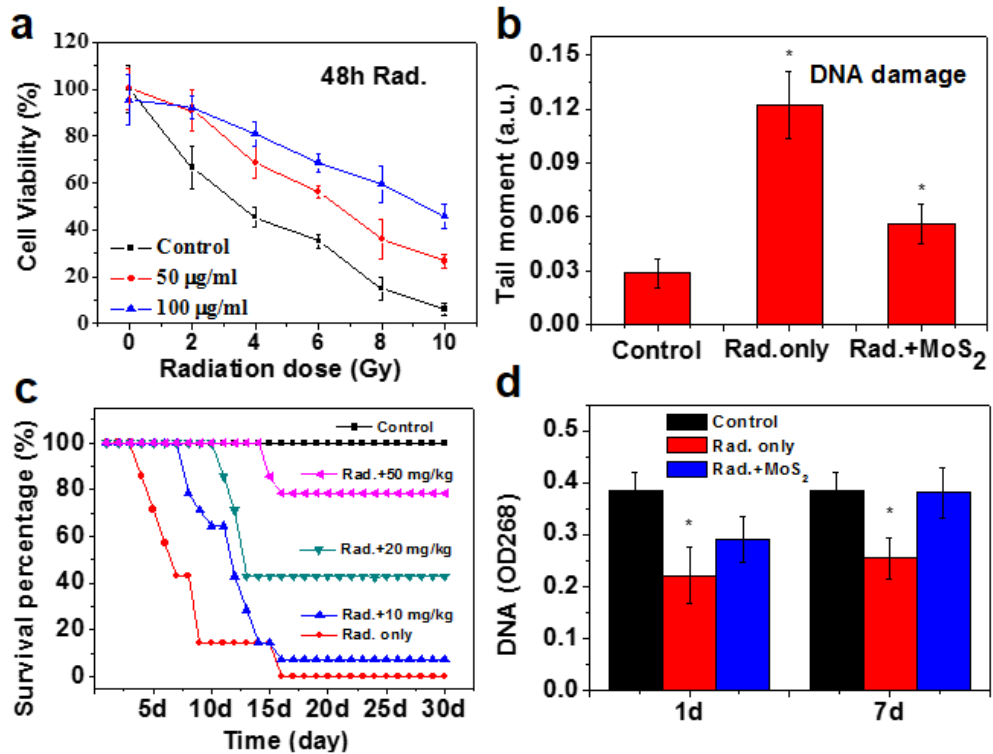

**Figure 2 Radiation protection *in vitro* and *in vivo*.** (**a**) Radiation dose-dependent protection i*n vitro* with different injected doses (50 and 100 μg/mL) or without treatments of cysteine-protected $MoS_2$ dots. The survival rates of A31 cells with treatments of cysteine-protected $MoS_2$ dots increased with increasing radiation doses, implying a significant function of protection against ionizing radiation *in vitro*. (**b**) Cell tail moment of mice with or without treatments of cysteine-protected $MoS_2$ dots, suggesting of its role in DNA repairs. Tail moment of DNA fragments from single cells receiving 4 Gy radiation was quantitatively calculated by counting 100 individual healthy and irradiated cells respectively and significant DNA recovery could be observed after being treated with cysteine-protected $MoS_2$ dots. (**c**) Survival curves of mice with different doses (10, 20 and 50 mg/kg) or without treatments of cysteine-protected $MoS_2$ dots (*n*=14 mice/group) showing an overall 79% survival proportion after being treated with a dose of 5 mg/ml. (**d**) DNA damages of mice 1 and 7 days after treatments of cysteine-protected $MoS_2$

dots (*n*=14 mice/group), as measured by UV-vis absorption at 268 nm. *P<0.05 as compared with the control group (paired Student's t-test).

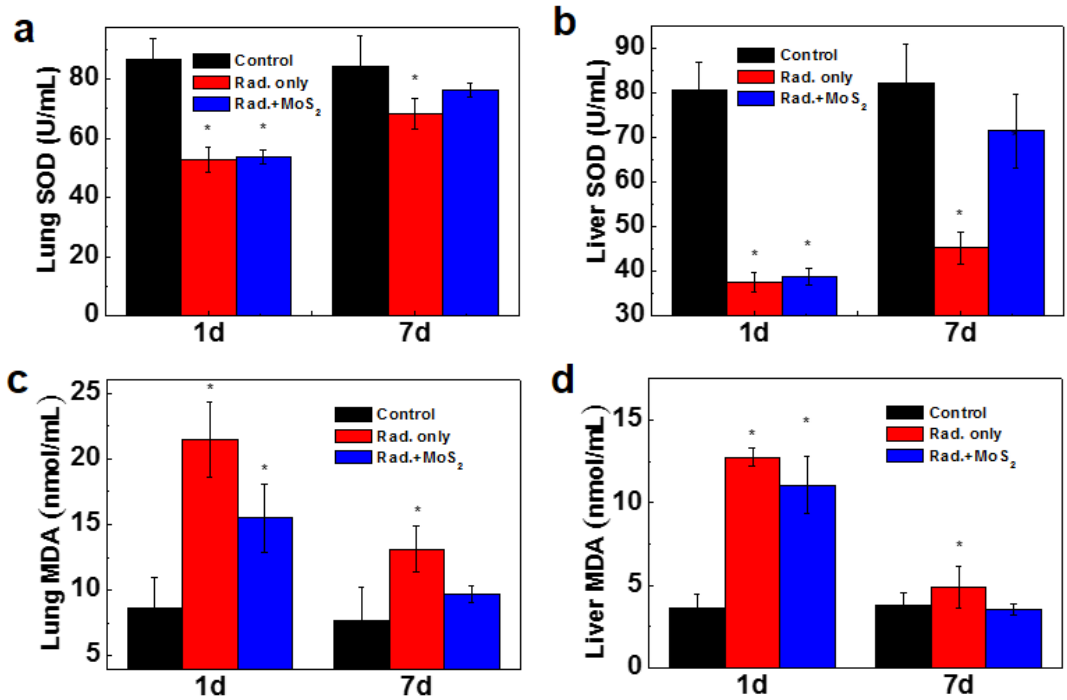

**Figure 3 Mechanisms of protection against radiation.** Superoxide dismutase (SOD) levels in (**a**) lung and (**b**) liver 1 and 7 days after gamma radiation with or without treatments of cysteine-protected $MoS_2$ dots ($n$=10 mice/group). 3,4-Methylenedioxyamphetamine (MDA) levels in (**c**) lung and (**d**) liver 1 and 7 days after gamma radiation with or without treatments of cysteine-protected $MoS_2$ dots ($n$=8 mice/group). Both SOD and MDA of the lung and liver recovered to healthy levels 7 days after treatments. *$P<0.05$ as compared with the control group (Student's t-test).

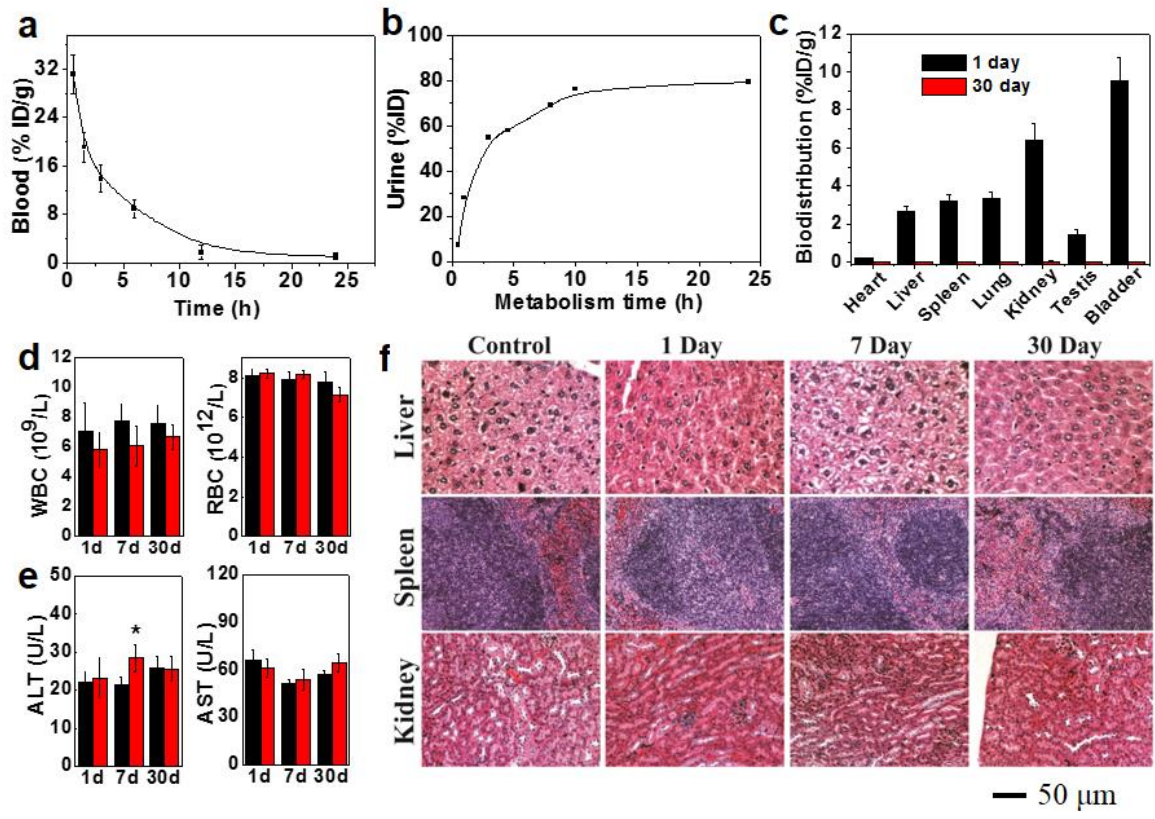

**Figure 4 *In vivo* pharmacokinetics, renal clearance, biodistribution and toxicities**. (**a**) Time-dependent concentrations in blood after intraperitoneal injection (**b**) Cumulative urine excretion at different time points. A large proportion of cysteine-protected $MoS_2$ dots in blood were cleared through the renal route. (**c**) Biodistribution of cysteine-protected $MoS_2$ dots at 24 hours and 30 days post injection. %ID/g=percentage of the injected dose per gram of weight. (**d**) Hematological data of RBC and WBC in mice treated with cysteine-protected $MoS_2$ dots. Data were collected at different time points of 1, 7 and 30 days after intraperitoneal injection (50 mg/kg). (**e**) Blood biochemistry analysis of mice treated with the cysteine-protected $MoS_2$ dots at different time points of 1, 7 and 30 days after treatments. *P<0.05 as compared with the control group (Student's t-test). (**f**) Pathological evaluation of liver, spleen and kidney of mice treated

with cysteine-protected MoS$_2$ dots. (*n*=8 mice/group) collect at different time points of 1, 7 and 30 days post injection. Scale bars, 50 μm.

**Supporting Information**

# Highly Catalytic Nanodots with Renal Clearance for Radiation Protection

*Xiao-Dong Zhang [1,3],\*, Jinxuan Zhang [2], Junying Wang [1], Jiang Yang [4], Jie Chen [3], Xiu Shen [3], Jiao Deng [5], Dehui Deng [5], Wei Long [3], Yuan-Ming Sun [3], Changlong Liu,[1] Meixian Li [2,] \**

[1] Department of Physics, School of Science, Tianjin University, Tianjin 300072, PR China

[2] Institute of Analytical Chemistry, College of Chemistry and Molecular Engineering, Peking University, Beijing 100871, China

[3] Tianjin Key Laboratory of Molecular Nuclear Medicine, Institute of Radiation Medicine, Chinese Academy of Medical Sciences and Peking Union Medical College, No. 238, Baidi Road, Tianjin 300192, China

[4] Environment, Energy and Natural Resources Center, Department of Environmental Science and Engineering, Fudan University, No.220, Handan Road, 200433, China

[5] State Key Laboratory of Catalysis, iChEM, Dalian Institute of Chemical Physics, Chinese Academy of Sciences, Dalian 116023, China

Correspondence should be addressed to X.Z. (xiaodongzhang@tju.edu.cn),

M.L.(lmwx@pku.edu.cn)

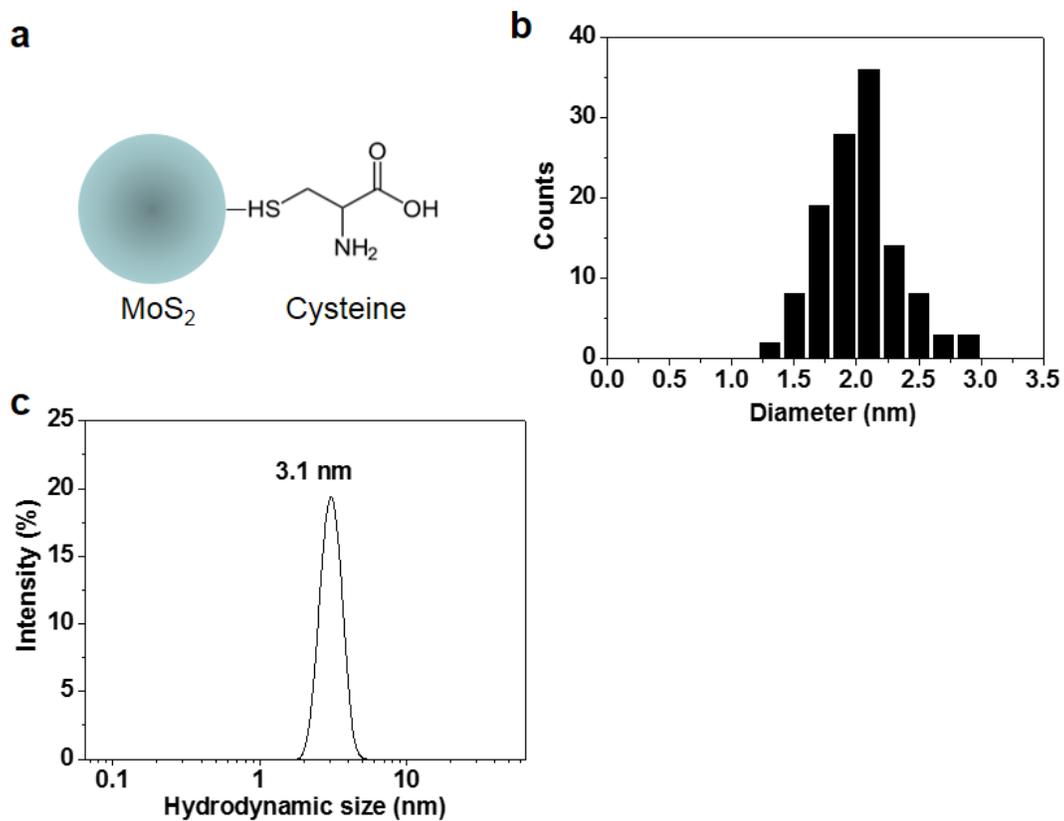

**Figure S1** (**a**) Schematic structure (not to scale) and (**b**) core size distribution of cysteine-protected $MoS_2$ dots measured by analyzing 121 nanodots from TEM images. (**c**) Hydrodynamic size of cysteine-protected $MoS_2$ dots measured by DLS.

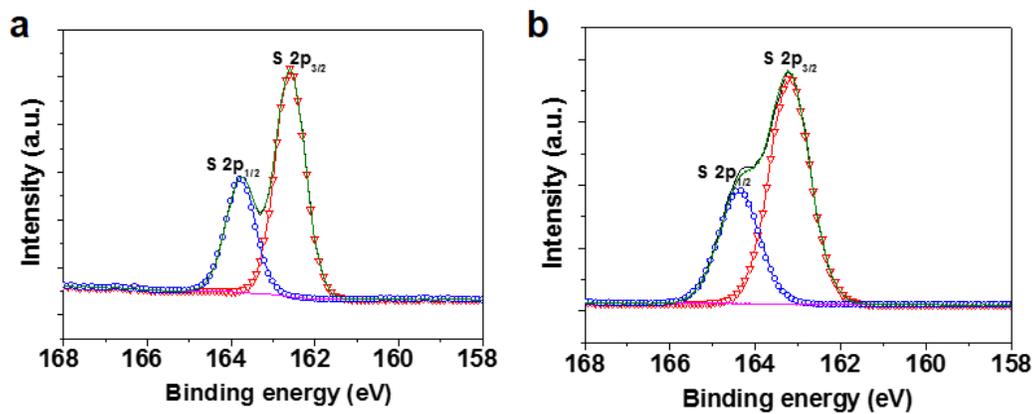

**Figure S2** XPS spectra of S 2p region for (**a**) unprotected MoS$_2$ dots and (**b**) cysteine. No disulfide states could be observed in both cases.

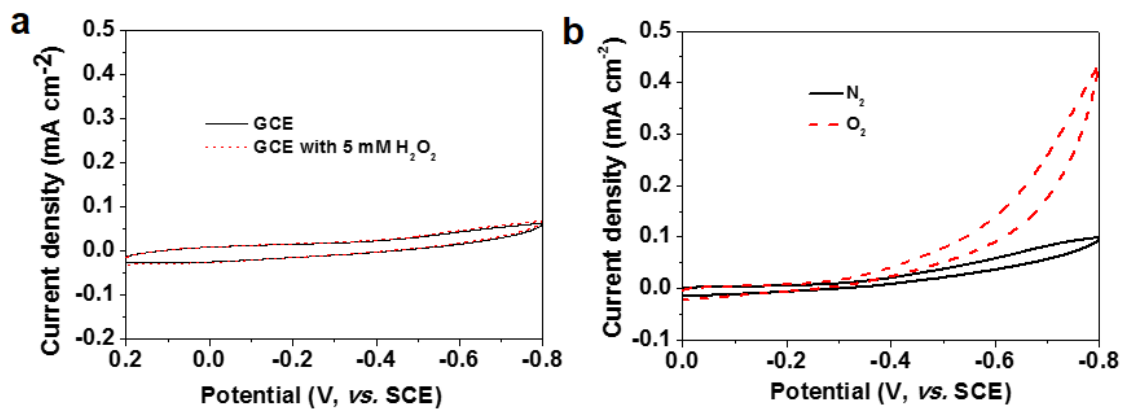

**Figure S3 Electrocatalytic activities of unmodified GCE**. (**a**) CVs of unmodified GCE with (dotted) or without (solid) addition of 5.00 mM $H_2O_2$ in $N_2$-saturated 0.01 M pH 7.4 PBS, (**b**) CVs of unmodified GCE in $N_2$- (solid) and $O_2$-saturated (dotted) 0.01 M pH 7.4 PBS. Scan rate: 50 mV s$^{-1}$. The electrocatalytic measurements indicate low reduction activities of unmodified GCE against $H_2O_2$ and $O_2$ without modification of cysteine protected-$MoS_2$ dots. Note that the as-shown current densities are in a much smaller scale than in **Figure. 1**.

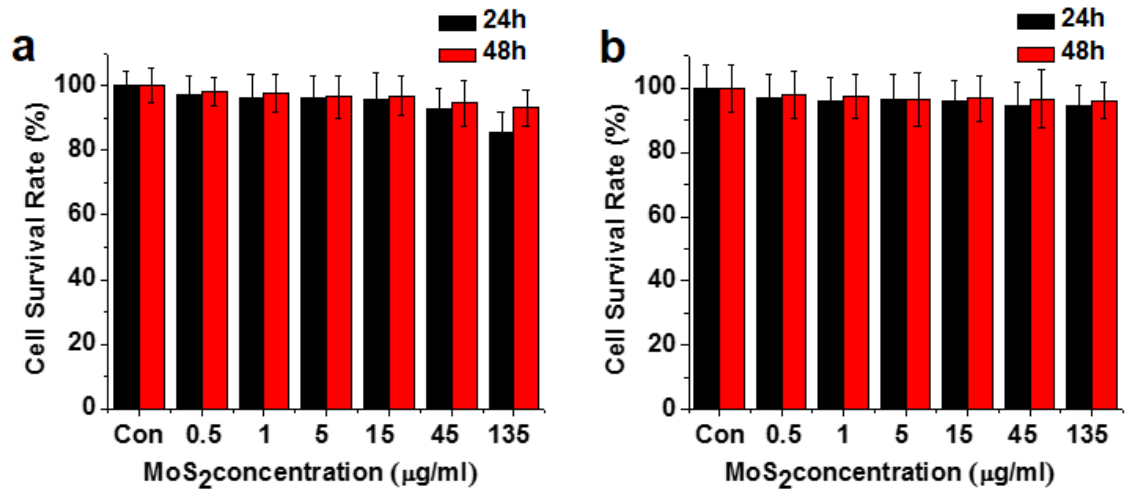

**Figure S4 Evaluation of cytotoxicities on cysteine-protected MoS$_2$ dots.** *In vitro* cytotoxicities of cysteine-protected MoS$_2$ dots on A31 mouse fibroblasts estimated by (**a**) Neutral Red assays and (**b**) MTT assays. The results were consistent and demonstrated low cytotoxicities of cysteine-protected MoS$_2$ dots.

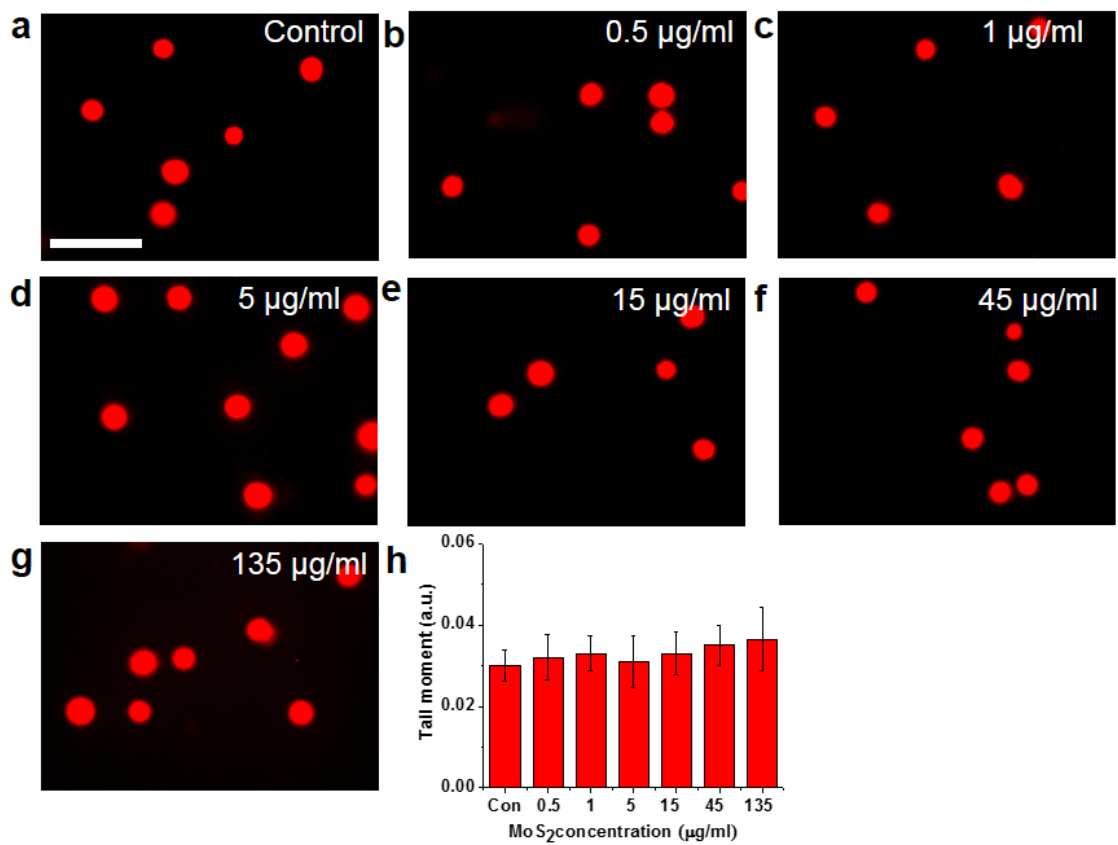

**Figure S5 Evaluation of DNA damages on cysteine-protected MoS₂ dots.** *In vitro* images of comet assays on A31 cells treated with different concentrations of MoS₂ dots (0.5-135 μg/ml) and **(h)** corresponding tail moment analysis.

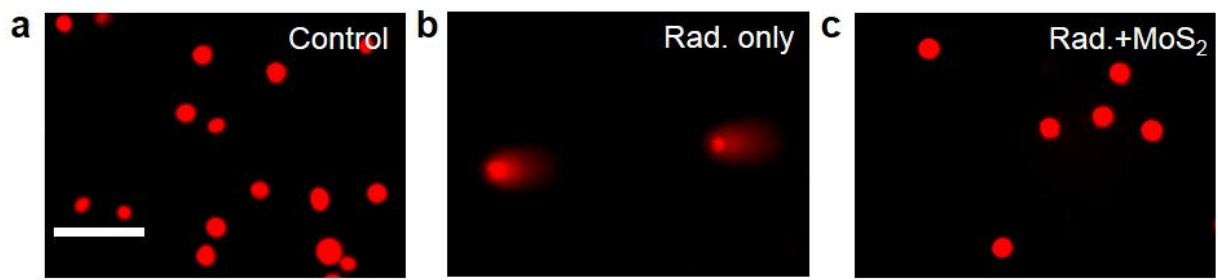

**Figure S6 (a-c)** Images of untreated control A31 cells, cells treated with only radiation and cysteine-protected MoS$_2$ dots plus radiation. The scale bar represent 100 μm.

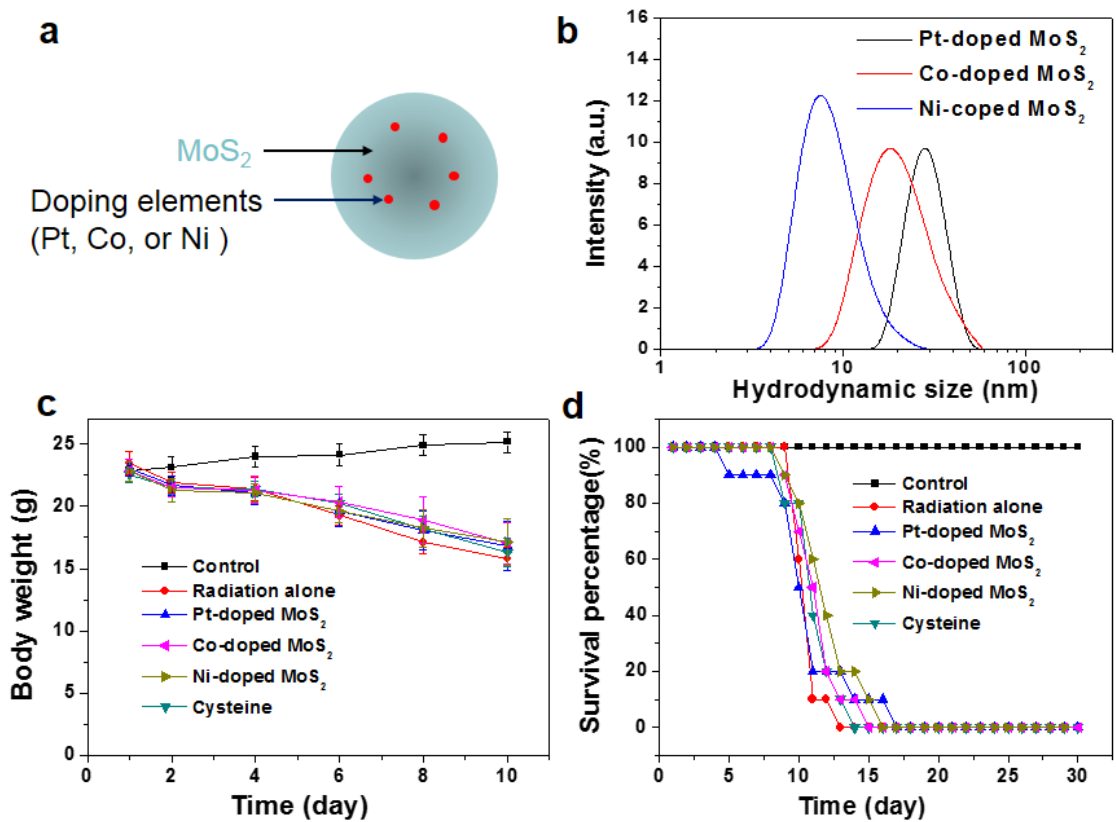

**Figure S7** (**a**) Schematic diagram showing Pt-, Co- or Ni-doped MoS$_2$ dots. (**b**) Hydrodynamic sizes of different metal-doped MoS$_2$ dots. (**c**) Body weight and (**d**) survival curves of mice treated with Pt-, Co-, Ni-doped MoS$_2$ dots and cysteine at a dose of 50 mg/kg under exposure to gamma ray. Compared with cysteine-protected MoS$_2$ dots, neither metal doping in MoS$_2$ dots nor pure cysteine showed increased survival proportions or reservation in body weight ($n$=10 mice/group).

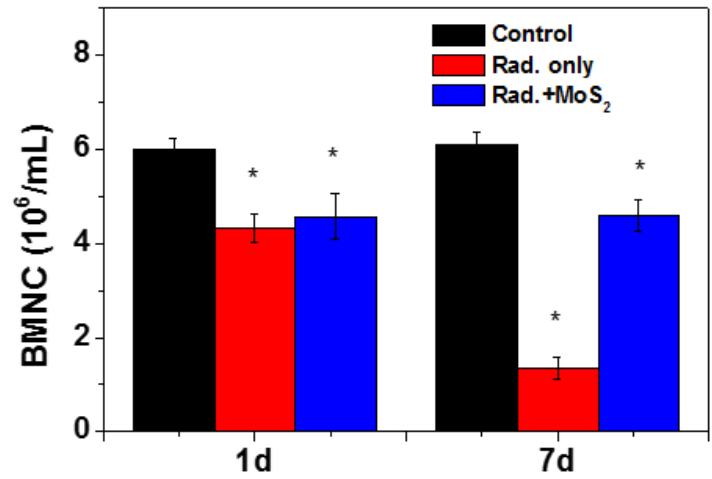

**Figure S8 Effects on BMNC.** Counts of bone marrow nucleated cells in control mice, mice only treated with radiation and mice treated with radiation and cysteine protected-$MoS_2$ dots at the dose of 50 mg/kg. Treatments of cysteine-protected $MoS_2$ dots almost restored BMNC was to normal levels. *$P<0.05$ as compared with the control group (Student's t-test).

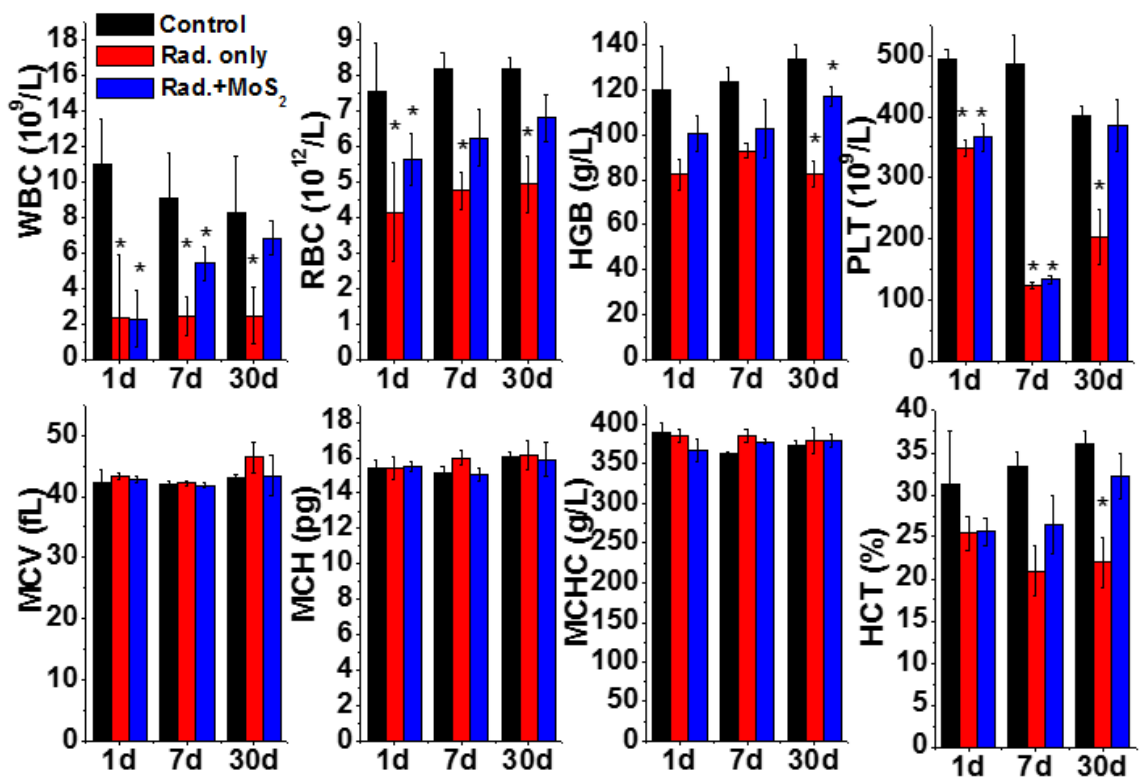

**Figure S9** *In vivo* **radiation protection of cysteine-protected MoS$_2$ dots**. (**a**) Hematological analysis of irradiated mice treated with or without cysteine-protected MoS$_2$ dots. Data were collected at different time points of 1, 7 and 30 days after intraperitoneal injection (50 mg/kg, *n*=10 mice/group). The results included red blood cells (RBC), white blood cells (WBC), platelets (PLT), mean corpuscular hemoglobin (MCH), mean corpuscular hemoglobin concentration (MCHC), mean corpuscular volume (MCV), hemoglobin (HGB), and hematocrit (HCT). Data were analyzed using Student's t-test and * in (a) and (b) indicates $p < 0.05$.

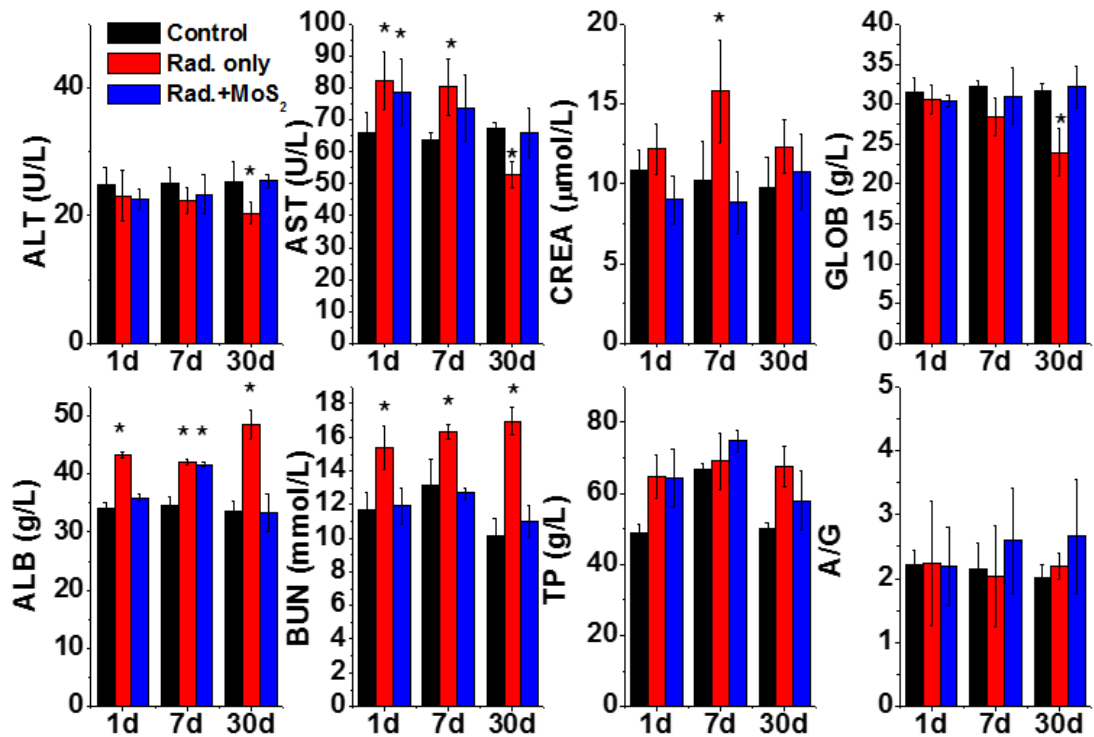

**Figure S10** *In vivo* radiation protection of cysteine-protected MoS$_2$ dots. (**b**) Blood biochemistry analysis of irradiated mice treated with or without cysteine-protected MoS$_2$ dots at different time points of 1, 7 and 30 days (50 mg/kg, *n*=10 mice/group). The results showed mean and standard deviation of alanine aminotransferase (ALT), aspartate aminotransferase (AST), total protein (TP), albumin (ALB), blood urea nitrogen (BUN), creatinine (CREA), globulin (GOLB) and total bilirubin (TBIL). The results from hematology and biochemistry panels showed the cysteine-protected MoS$_2$ dots can recover the indicators 30 days post injection. Data were analyzed using Student's t-test and * in (a) and (b) indicates p < 0.05.

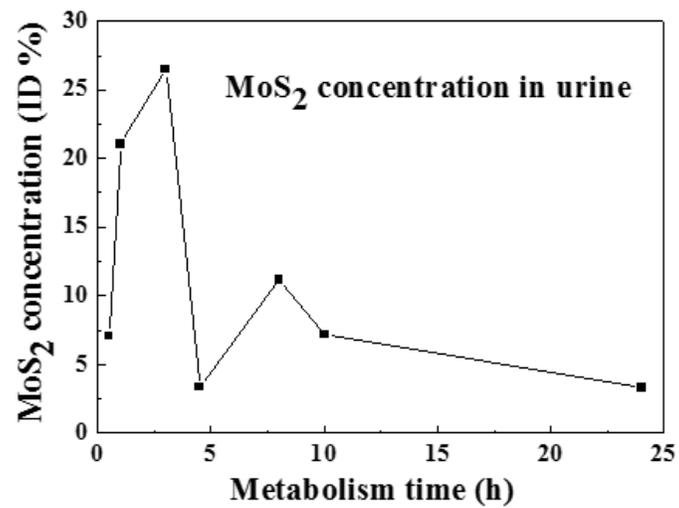

**Figure S11 Time-course renal clearance of cysteine-protected MoS$_2$ dots.** Time-dependent concentrations of cysteine-protected MoS$_2$ dots (50 mg/kg) in urine of mice at different time points post injection, measured by ICP-MS.

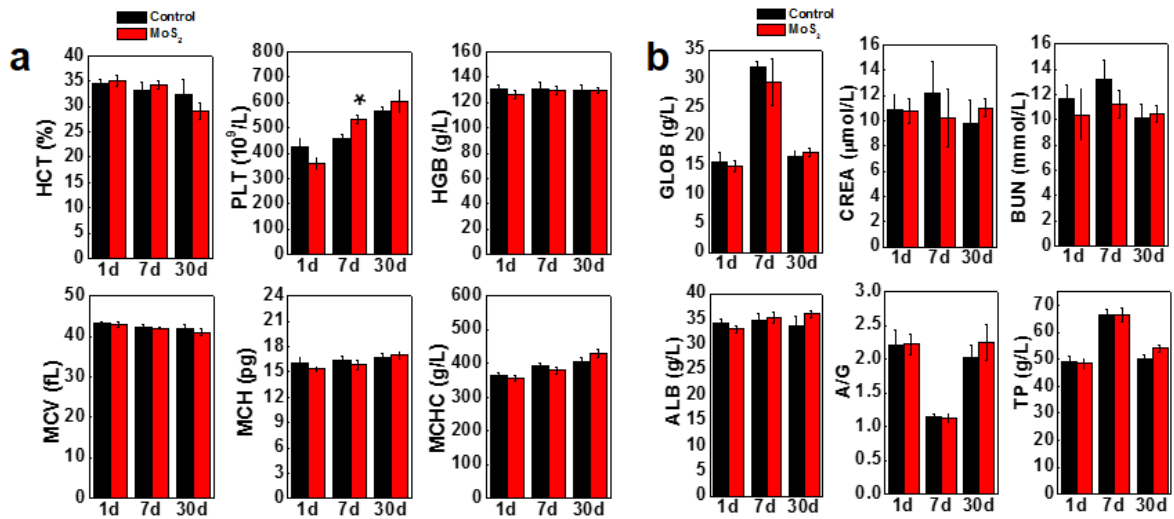

**Figure S12 *In vivo* toxicities of cysteine-protected MoS$_2$ dots**. (**a**) Hematological data of mice treated with cysteine protected-MoS$_2$ dots. Data were collected at different time points of 1, 7 and 30 days after intraperitoneal injection (50 mg/kg, *n*=8 mice/group). The results included PLT, MCH, MCHC, MCV, HGB, and HCT.(**b**) Blood biochemistry analysis on mice treated with cysteine-protected MoS$_2$ dots at different time points of 1, 7 and 30 days. The results showed mean and standard deviation of TP, ALB, BUN, CREA, GOLB, and TBIL. Data were analyzed using Student's t-test and * in (a) and (b) indicates p < 0.05.

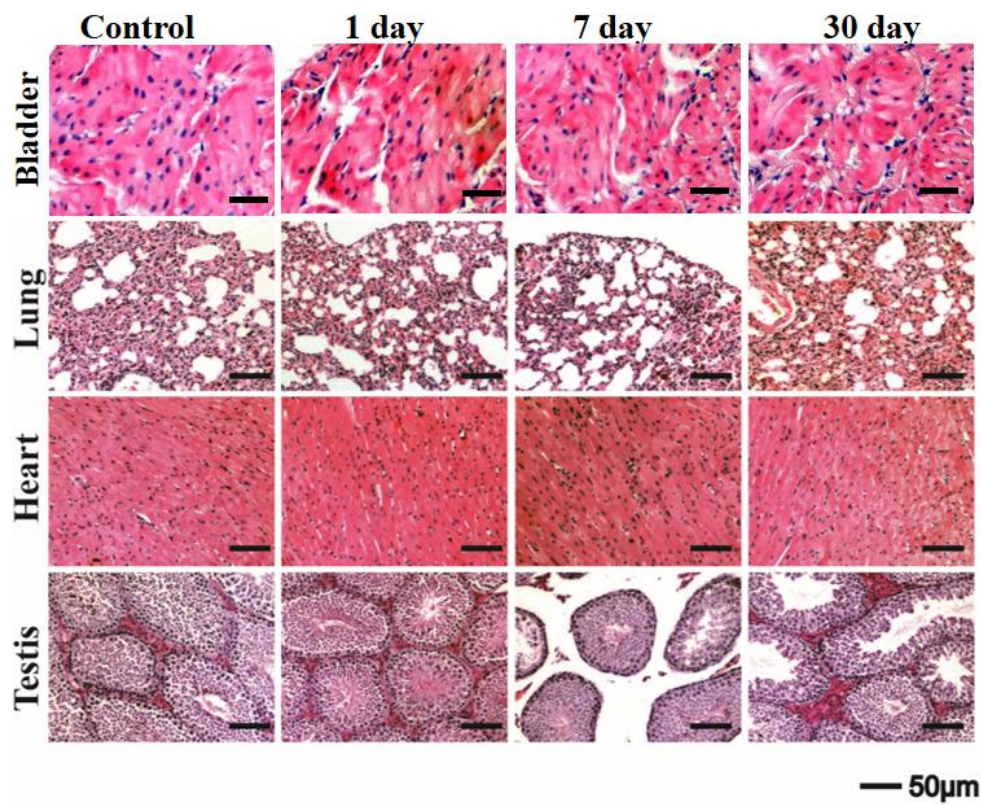

**Figure S13** Representative pathological analysis for bladder, lung, heart and testis of mice treated with cysteine-protected $MoS_2$ dots (50 mg/kg, *n*=8 mice/group) at different time points of 1, 7 and 30 days. Scale bars, 50 μm.